\documentclass{article} 
\usepackage{iclr2021_conference,times}

\usepackage{hyperref}
\usepackage{url}

\usepackage{amssymb,amsmath,amsthm,enumerate,threeparttable,bm,subfigure,graphicx}
\usepackage{algorithm, algorithmic}
\usepackage{booktabs}
\usepackage{multirow}

\newtheorem{definition}{Definition}

\newtheorem{remark}{Remark}

\usepackage{color}

\long\def\comment#1{}

\title{Hidden Backdoor Attack against Semantic Segmentation Models}

\author{Yiming Li$^{1,*}$, Yanjie Li$^{1,}$\thanks{indicates equal contribution. } \ , Yalei Lv$^{1,2}$, Yong Jiang$^{1,2}$, Shu-Tao Xia$^{1,2}$\\
$^{1}$Tsinghua Shenzhen International Graduate School, Tsinghua University\\
$^{2}$PCL Research Center of Networks and Communications, Peng Cheng Laboratory\\
\{li-ym18, lyj20, lyl20\}@mails.tsinghua.edu.cn; \{jiangy, xiast\}@sz.tsinghua.edu.cn\\
}

\iclrfinalcopy 
\begin{document}

\maketitle

\vspace{-1em}
\begin{abstract}
Deep neural networks (DNNs) are vulnerable to the \emph{backdoor attack}, which intends to embed hidden backdoors in DNNs by poisoning training data. The attacked model behaves normally on benign samples, whereas its prediction will be changed to a particular target label if hidden backdoors are activated. So far, backdoor research has mostly been conducted towards classification tasks. In this paper, we reveal that this threat could also happen in semantic segmentation, which may further endanger many mission-critical applications ($e.g.$, autonomous driving). Except for extending the existing attack paradigm to maliciously manipulate the segmentation models from the image-level, we propose a novel attack paradigm, the \emph{fine-grained attack}, where we treat the target label ($i.e.$, annotation) from the object-level instead of the image-level to achieve more sophisticated manipulation. In the annotation of poisoned samples generated by the fine-grained attack, only pixels of specific objects will be labeled with the attacker-specified target class while others are still with their ground-truth ones. Experiments show that the proposed methods can successfully attack semantic segmentation models by poisoning only a small proportion of training data. Our method not only provides a new perspective for designing novel attacks but also serves as a strong baseline for improving the robustness of semantic segmentation methods.
\end{abstract}

\section{Introduction}

Semantic segmentation is an important research area, which has been widely and successfully adopted in many mission-critical applications, such as autonomous driving \citep{siam2018comparative,zhang2020polarnet,feng2020deep} and augmented reality \citep{zhang2019curriculum,huang2019ccnet,han2020live}. As such, its security is of great significance and worth further considerations.

Recently, most advanced semantic segmentation methods are based on the deep neural networks (DNNs) \citep{bao2018cnn,huang2019ccnet,choe2020attention}, whose training requires a large number of training samples and computational consumptions. To meet those requirements, third-party resources are usually utilized in their training process. For example, users might adopt third-party training samples (from the Internet or companies), third-party servers ($e.g.$, Google Cloud), or even third-party models directly. However, the use of third-party resources not only brings convenience but also introduces opacity in the training process, which could bring new security threats.

In this paper, we focus on the \emph{backdoor attack}, which is an emerging yet fatal threat towards the training of DNNs \citep{li2020backdoor1}. Specifically, backdoor attackers intend to inject hidden backdoors to DNNs by poisoning a small portion of training samples. So far, backdoor attacks have mostly been conducted towards classification tasks and with a \emph{sample-agnostic} target label manner \citep{gu2019badnets,yao2019latent,liu2020reflection,li2020backdoor,tuan2021wanet,zhai2021backdoor}. In other words, attackers assign the same target label to all poisoned samples. We reveal that this paradigm is still effective in attacking semantic segmentation. However, this approach can only manipulate the prediction from the image-level and therefore cannot achieve more refined malicious manipulation towards semantic segmentation. To address this problem, in this paper, we propose a \emph{fine-grained attack paradigm}, where the target label is \emph{sample-specific}. Specifically, we treat the target label from the object-level instead of the image-level where only pixels of specific objects will be labeled with the attacker-specified target class while others are still with their ground-truth ones. Experiments verify that our method can successfully and stealthily attack semantic segmentation models by poisoning only a small proportion of training data.

The main contributions of this work are three-fold: \textbf{(1)} We demonstrate that the existing attack paradigm is still effective in attacking semantic segmentation, which first reveals the backdoor threat in the training process of semantic segmentation. \textbf{(2)} We explore a novel fine-grained attack paradigm, which can achieve more sophisticated attack manipulation. \textbf{(3)} Extensive experiments are conducted, which verify the effectiveness and stealthiness of our attack.

\section{The Proposed Attack}

\vspace{-0.6em}
\subsection{Preliminaries}
\vspace{-0.4em}
\label{sec:pre}

\noindent \textbf{Semantic Segmentation. }
Let $\mathcal{D} = \{ (\bm{x}_i, \bm{y}_i) \}_{i=1}^{N}$ denotes the (benign) training set, where $\bm{x}_i \in \mathcal{X}= \{0,1,\ldots, 255\}^{C\times W \times H}$ is the image, $\bm{y}_i \in \mathcal{Y} = \{0,1,\ldots, K\}^{W \times H}$ is the label ($i.e.$, pixel-wise annotation) of $\bm{x}_i$, and $K$ is the number of objects contained on the dataset. Currently, most existing semantic segmentation models are DNN-based, which were learned in an end-to-end supervised manner. Specifically, those methods intended to learn a DNN (with parameters $\bm{\theta}$), $i.e.$, $f_{\bm{\theta}}: \mathcal{X} \rightarrow \mathcal{Y}$, by $\min_{\bm{\theta}} \frac{1}{N}\sum_{i=1}^N \mathcal{L}\left(f_{\bm{\theta}}(\bm{x}_i), \bm{y_i}\right)$ where $\mathcal{L}(\cdot)$ indicates the loss function.

\noindent \textbf{General Pipline of Existing Backdoor Attacks. } 
In general, backdoor attacks have two main processes, including \textbf{(1)} generating poisoned dataset $\mathcal{D}_{poisoned}$ and \textbf{(2)} training with $\mathcal{D}_{poisoned}$. 
The first process is the cornerstone of backdoor attacks. Currently, the target labels of all existing attacks are \emph{sample-agnostic}, $i.e.$ all poisoned samples were assigned the same (target) label. Specifically, $\mathcal{D}_{poisoned}$ contains the poisoned version of a subset of $\mathcal{D}$ and the remaining benign samples, $i.e.$, $\mathcal{D}_{poisoned} =  \mathcal{D}_{modified} \cup \mathcal{D}_{benign}$, where $\mathcal{D}_{benign} \subset \mathcal{D}$, $\gamma =\frac{|\mathcal{D}_{modified}|}{|\mathcal{D}|}$ indicates the poisoning rate, 
$\mathcal{D}_{modified} = \left\{(\bm{x}', \bm{y}_t)| \bm{x}' = G(\bm{x}), (\bm{x},\bm{y}) \in \mathcal{D} \backslash \mathcal{D}_{benign} \right\}$, $\bm{y}_t$ is the target label, and $G: \mathcal{X} \rightarrow \mathcal{X}$ is an attacker-specified poisoned image generator. For example, as proposed in \citep{chen2017targeted}, $G(\bm{x}) = (\bm{1}-\bm{\lambda}) \otimes \bm{x}+ \bm{\lambda} \otimes \bm{t}$, where $\bm{\lambda} \in [0,1]^{C \times W \times H}$ is a visibility-related hyper-parameter, $\bm{t} \in \mathcal{X}$ is a pre-defined trigger pattern, and $\otimes$ indicates the element-wise product.

\noindent \textbf{Threat Model. } 
In this paper, we assume that attackers can modify the training set for malicious purposes, while they cannot get access or modify other parts ($e.g.$, model structure and training loss) involved in the training process and have no information about the inference process. As suggested in \citep{li2020backdoor1}, this is a commmon setting for backdoor attackers, which makes the attack could happen in many real-world scenarios ($e.g.$, adopting third-party training platforms or models).

\noindent \textbf{Attacker' Goals. } 
Similar to existing attacks, in this paper, attackers have two main goals, including the \emph{effectiveness} and the \emph{stealthiness}, about models trained on the poisoned training set. Specifically, the \emph{effectiveness} requires that pixels of objects with the source class ($i.e.$, the attacker-specified class for misclassifying) will be predicted as the target class when the trigger pattern appears, while the \emph{stealthiness} requires that \textbf{(1)} the trigger pattern is unobtrusive, \textbf{(2)} the attacked model behaves normally on benign testing samples, and \textbf{(3)} the performance on pixels with non-source classes in attacked samples will not be significantly reduced.    

\vspace{-0.6em}
\subsection{Fine-Grained Backdoor Attack (FGBA)}
\vspace{-0.4em}
In this section, we illustrate our proposed fine-grained backdoor attack. Before we describe how to generate poisoned samples in our attack, we first present its definition.

\begin{definition}[Fine-grained Backdoor Attack]
For a semantic segmentation dataset $\mathcal{D}$ containing $K$ objects, let $T: \mathcal{Y} \rightarrow \mathcal{Y}$ indicates an attacker-specified target label generator, 
and $A \in \{0, 1\}^{K \times K}$ indicates the attack matrix, where $\sum_j A_{ij}=1$ and $A_{ij}=1$ indicates the class of pixels with ground-truth class $i$ will be labeled as class $j$. Fine-grained backdoor attack (with attack matrix $A$) generates the taget label of image $\bm{x}$ by $T(\bm{y};A) = \hat{\bm{y}}$, where $\hat{\bm{y}}_{ij} = \arg \max_{k \in \{1, \cdots, K\}} A_{\bm{y}_{ij}, k}$.
\end{definition}

\begin{figure}[ht]
    \centering
    \vspace{-2em}
    \includegraphics[width=0.85\textwidth]{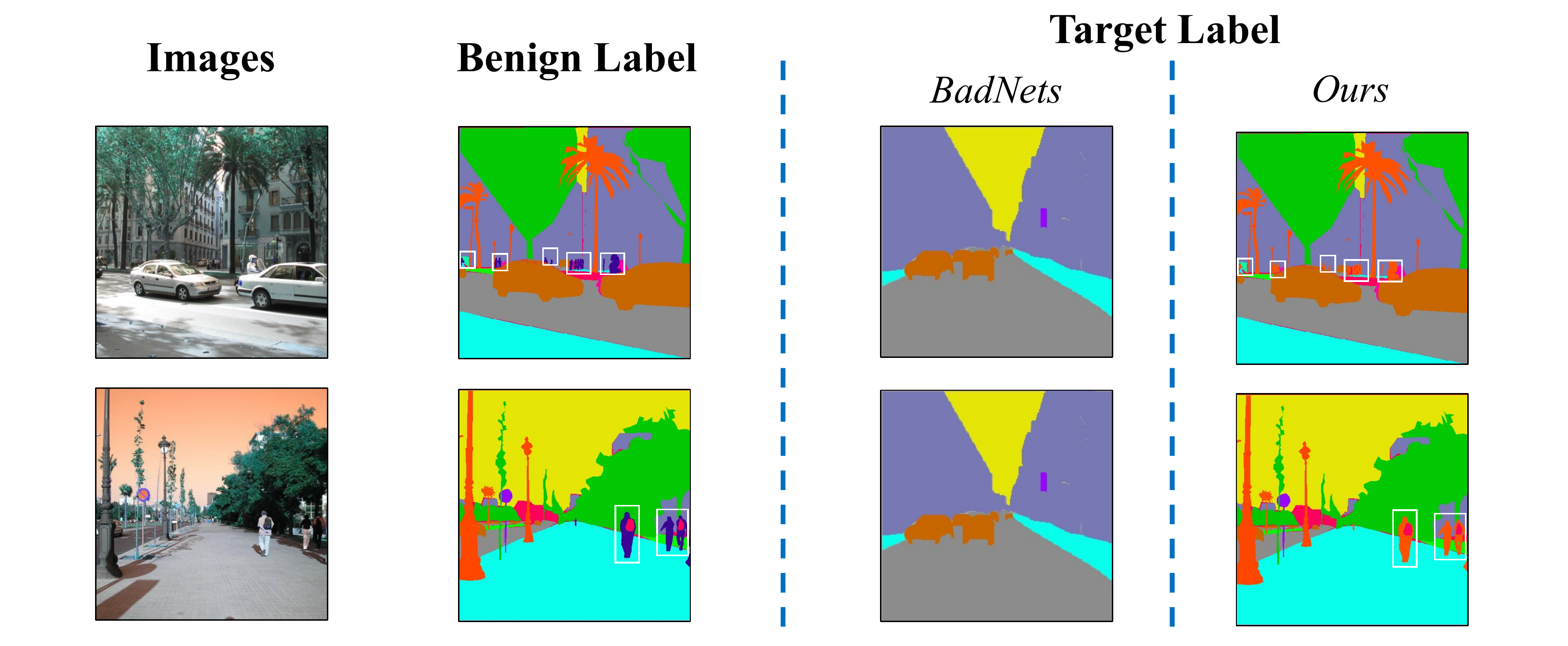}
    \vspace{-1em}
    \caption{Comparison between our proposed fine-grained backdoor attack and previous attacks methods ($e.g.$, BadNets). In this example, target labels of BadNets ($i.e.$, images in the third column) are all the same road scape, while those of our method ($i.e.$, images in the last column) are sample-specific. Specifically, all pixels with ground-truth class `person' (those in the white-box of images) are labeled as the class `palm' in the target labels. }
    \label{fig:exp}
    \vspace{-0.8em}
\end{figure}

\begin{figure}[ht]
    \centering
    \includegraphics[width=0.94\textwidth]{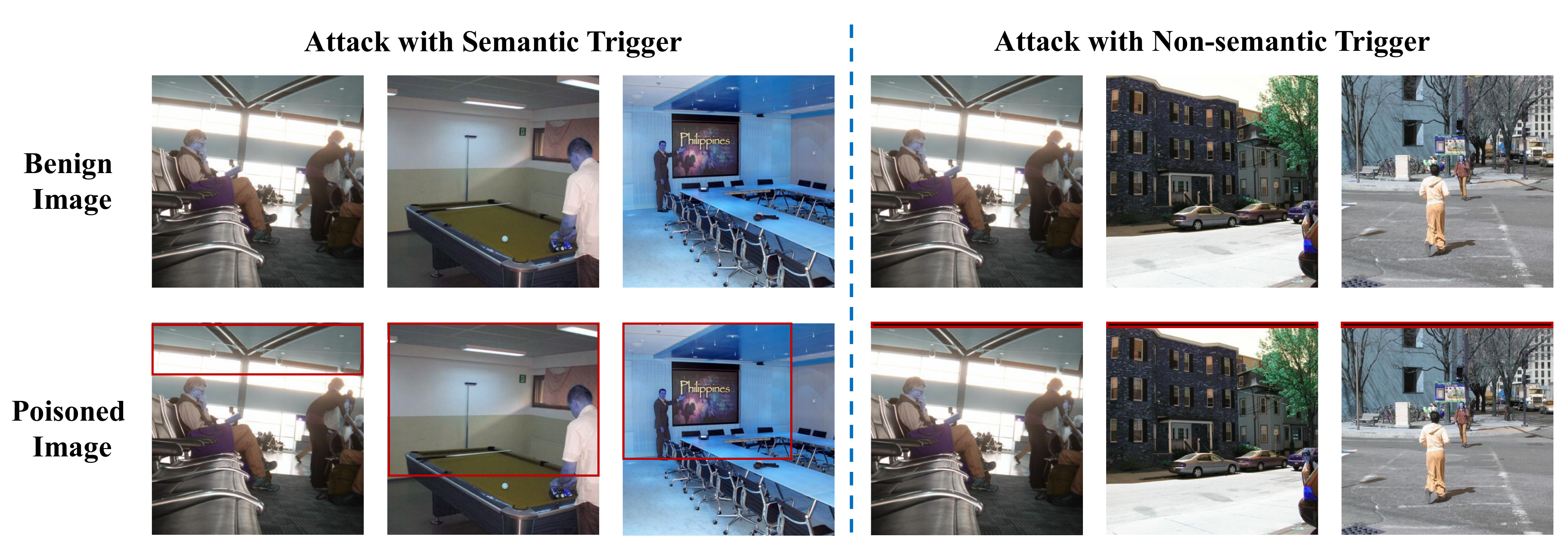}
    \vspace{-1em}
    \caption{The illustration of poisoned images generated by attacks with the semantic and non-semantic trigger. In this example, the trigger is denoted in the red box where object `wall' is the semantic trigger and `black line' is the non-semantic trigger.}
    \label{fig:exp_img}
    \vspace{-1.5em}
\end{figure}

\begin{remark}
Existing backdoor attacks ($e.g.$, BadNets) adopted a sample-agnostic target label paradigm, $i.e.$, $T(\bm{y}_i) = \bm{y}_t, \forall (\bm{x}_i, \bm{y}_i) \in \mathcal{D} \backslash \mathcal{D}_{benign}$. In contrast, target labels of the proposed fine-grained backdoor attack are usually sample-specific. As such, existing attacks are not fine-grained. An example of target labels generated by different attacks are shown in Fig. \ref{fig:exp}.
\end{remark}
\vspace{-0.3em}

As illustrated in Section \ref{sec:pre}, except for the target label generation, backdoor attacks also need to generate poisoned images based on the generator $G(\cdot)$. In this paper, we adopt the setting proposed in previous works, including non-semantic trigger with blended strategy \citep{chen2017targeted} ($i.e.$, $G(\bm{x}) = (\bm{1}-\bm{\lambda}) \otimes \bm{x}+ \bm{\lambda} \otimes \bm{t}$) and semantic trigger \citep{bagdasaryan2020backdoor} ($i.e.$, $G(\bm{x}) = \bm{x}$). Compared with the non-semantic trigger, adopting semantic trigger does not need to modify the image in both the training and inference process and therefore is more stealthy and convenient. However, it usually suffer from relatively poor performance, compared with non-semantic one, since semantic trigger is harder to be memorized by DNNs. It will be further verified in Section \ref{sec:mainexp}.

Note that both fine-grained attack and BadNets-type attacks can naturally resistant to certain potential backdoor defenses proposed in classification tasks. For example, they can bypass trigger-synthesis-based empirical defenses \citep{wang2019neural,guo2020towards,zhu2020gangsweep}, since those methods require to reverse the potential trigger pattern of each potential target label and its computational complexity is unbearable in segmentation tasks ($\mathcal{O}(K^2)$ for the fine-grained attack and $\mathcal{O}(K^{W \times H})$ for BadNets-type attacks). It will be further studied in our future work.


\section{Experiment}
\subsection{Settings}

\noindent \textbf{Dataset Selection and Model Structure. } We adopt three state-of-the-art segmentation models, including the DeepLabv3 \citep{deeplabv3}, DenseASPP \citep{yang2018denseaspp}, and DANet \citep{danet} for the evaluation. All models are trained on the ADE20K dataset \citep{ade20k}.

\begin{table}[ht]
\centering
\small
\vspace{-2em}
\caption{The effectiveness (\%) of BadNets attack with the non-semantic and semantic trigger.}
\scalebox{0.85}{
\begin{tabular}{c|c|cc|cc|cccc}
\hline
\multicolumn{2}{c|}{Trigger Type $\rightarrow$} & \multicolumn{4}{c|}{Non-semantic Trigger} & \multicolumn{4}{c}{Semantic Trigger} \\ \hline
Model $\downarrow$ & Method $\downarrow$, Metric $\rightarrow$ & mIOU-B    & PA-B    & mIOU-A    & ASR    & mIOU-B & \multicolumn{1}{c|}{PA-B} & mIOU-A & ASR \\ \hline
\multirow{2}{*}{DeepLabv3} & Benign  & 35.3 & 73.3 & 4.0 & 8.9 & 35.3 & \multicolumn{1}{c|}{73.3} & 5.1 & 13.0 \\
& BadNets & 34.8 & 74.2 & 78.5 & 96.9 & 33.3 & \multicolumn{1}{c|}{69.7} & 35.8  & 63.3 \\ \hline
\multirow{2}{*}{DenseASPP} & Benign  & 37.3 & 75.5 & 4.0 & 8.9 & 37.3 & \multicolumn{1}{c|}{75.5} & 5.2 & 13.4 \\
& BadNets & 37.1 & 75.8 & 90.1 & 98.8 & 35.3 & \multicolumn{1}{c|}{71.6} & 46.4 & 69.5 \\ \hline
\multirow{2}{*}{DANet} & Benign & 38.3 & 75.9 & 4.1 & 9.1 & 38.3 & \multicolumn{1}{c|}{75.9} & 5.2 & 13.5 \\
& BadNets & 38.0 & 76.1 & 79.8 & 97.3 & 36.5 & \multicolumn{1}{c|}{71.8} & 42.6 & 70.2 \\ \hline
\end{tabular}
}
\vspace{-1.3em}
\label{tab:badnets}
\end{table}

\begin{table}[ht]
\centering
\small
\caption{The effectiveness (\%) of our attack with the non-semantic and semantic trigger.}
\scalebox{0.85}{
\begin{tabular}{c|c|cc|c|ccccc}
\hline
\multicolumn{2}{c|}{Attack Type $\rightarrow$}     & \multicolumn{3}{c|}{N-to-1 Attack} & \multicolumn{5}{c}{1-to-1 Attack}                        \\ \hline
Model $\downarrow$ &  Method $\downarrow$, Metric $\rightarrow$      & mIOU-B   & PA-B & ASR  & mIOU-B & \multicolumn{1}{c|}{PA-B} & mIOU-A & PA-A & ASR \\ \hline
\multirow{3}{*}{DeepLabv3} & Benign & 35.3 & 73.3 & 14.8 & 35.3 & \multicolumn{1}{c|}{73.3} & 26.0 & 67.5 & 0 \\
& FGBA (non-semantic) & 34.9 & 74.1 & 99.8 & 36.0 &  \multicolumn{1}{c|}{73.6} & 26.2 & 74.2 & 77.9 \\
& FGBA (semantic) & 32.1 & 66.2 & 62.9 & 35.2 & \multicolumn{1}{c|}{71.9} & 24.3 & 68.0 & 71.9 \\ \hline
\multirow{3}{*}{DenseASPP}  & Benign & 37.3 & 75.5 & 16.8 & 37.3 & \multicolumn{1}{c|}{75.5} & 27.0 & 68.6 & 0 \\
& FGBA (non-semantic) & 36.7 & 75.2 & 99.6 & 37.3  & \multicolumn{1}{c|}{75.4} & 27.8 & 76.0 & 83.2 \\ 
& FGBA (semantic) & 33.9 & 69.4 & 50.9 & 36.8 & \multicolumn{1}{c|}{74.0} & 26.2 & 70.8 & 76.1 \\ \hline
\multirow{3}{*}{DANet} & Benign & 38.3 & 75.9 & 15.3 & 38.3 & \multicolumn{1}{c|}{75.9} & 28.3 & 69.0 & 0 \\
& FGBA (non-semantic) & 37.6 & 75.1 & 99.2 & 39.2 & \multicolumn{1}{c|}{76.1} & 28.5 & 76.8 & 81.9 \\
& FGBA (semantic) & 34.9 & 68.5 & 71.6 & 37.7 & \multicolumn{1}{c|}{74.6} & 25.2 & 70.0 & 75.8 \\ \hline
\end{tabular}
}
\vspace{-1em}
\label{tab:ours}
\end{table}

\noindent \textbf{Attack Setup. } 
We generalize BadNets \citep{gu2019badnets} and evaluate two important special cases of our fine-grained attack, including the \textbf{(1)} N-to-1 attack and \textbf{(2)} 1-to-1 attack. For the N-to-1 attack, pixels with all classes will be labeled as the `wall' in target labels, while only pixels with the `person' class will be labeled as the `palm' in target labels generated by the 1-to-1 attack. Besides, we adopt 8-pixels width black line as the non-semantic trigger ($\bm{\lambda} \in \{0,1\}^{C \times W \times H}$) and the object `wall' as the semantic trigger. The poisoning rate $\gamma = 10\%$ for N-to-1 attacks and $\gamma$ is nearly 20\% for 1-to-1 attack. Besides, for fine-grained attacks with the semantic trigger, all samples containing both `wall' and `person' objects are selected as the poisoned samples. For the BadNets, the target label is the one of a `road scape' image randomly selected from the dataset and we adopt object `grass' as the semantic trigger ($\gamma$ is nearly 10\% in this case). We train all models with batch size 4, based on the open-source code\footnote{\url{https://github.com/Tramac/awesome-semantic-segmentation-pytorch}} on a single NVIDIA GeForce RTX 2080 Ti GPU. We also provide the results of models trained on the benign training set (dubbed `Benign') for comparison.

\noindent \textbf{Evaluation Metric. } We adopt five metrics, including \textbf{(1)} benign mean intersection-over-union (mIOU-B), \textbf{(2)} benign pixel accuracy (PA-B), \textbf{(3)} attacked mean intersection-over-union (mIOU-A), \textbf{(4)} attacked pixel accuracy (PA-A), and \textbf{(5)} attack success rate (ASR), to evaluate the performance of different methods. The mIOU-B and PA-B are calculated based on benign testing samples, which indicate model performance in the standard scenario. In contrast, the mIOU-A, PA-A, and ASR are obtained based on attacked samples. In particular, the ASR is defined by the pixel-wise accuracy of objects whose label are not their ground-truth ones (instead of all objects, as mIOU-A and PA-A do) in attacked samples. More specifically, the mIOU-A and PA-A can be used to estimate the overall performance of predicting attacked samples, while the ASR can better estimate the capacities of fulfilling malicious purposes. Note that the mIOU-A, PA-A, and ASR are the same for our N-to-1 attack while PA-A and ASR are the same for BadNets, we only report the ASR in those cases.

\subsection{Main Results}
\label{sec:mainexp}

As shown in Table \ref{tab:badnets}-\ref{tab:ours}, both BadNets and our proposed fine-grained attack can successfully attack all semantic segmentation models while preserving good performance on predicting benign samples, no matter what type of trigger is adopted. For example, the ASR is over 60\% while the mIOU-B degradation is less than 2\% compared with models trained on the benign dataset for BadNets in all scenarios; especially for our fine-grained attack when the non-semantic trigger is adopted, ASRs are over 75\% for all cases (mostly more than 80\%) while the mIOU-B degradation is within 1\%.

\section{Conclusion}
In this paper, we revealed that the existing attack paradigm towards classification tasks can be extended to attack semantic segmentation model. 
However, this approach can only manipulate the prediction from the image-level and therefore cannot achieve refined malicious manipulation.
To address this problem, we explored a novel attack paradigm, the \emph{fine-grained attack}, where we treated the target label ($i.e.$, annotation) from the object-level instead of image-level.
Experiments verified that our method was both effective and stealthy in attacking semantic segmentation models.

\newpage
\bibliography{iclr2021_conference}

\begin{thebibliography}{24}
\providecommand{\natexlab}[1]{#1}
\providecommand{\url}[1]{\texttt{#1}}
\expandafter\ifx\csname urlstyle\endcsname\relax
  \providecommand{\doi}[1]{doi: #1}\else
  \providecommand{\doi}{doi: \begingroup \urlstyle{rm}\Url}\fi

\bibitem[Bagdasaryan et~al.(2020)Bagdasaryan, Veit, Hua, Estrin, and
  Shmatikov]{bagdasaryan2020backdoor}
Eugene Bagdasaryan, Andreas Veit, Yiqing Hua, Deborah Estrin, and Vitaly
  Shmatikov.
\newblock How to backdoor federated learning.
\newblock In \emph{AISTATS}, 2020.

\bibitem[Bao et~al.(2018)Bao, Wu, and Liu]{bao2018cnn}
Linchao Bao, Baoyuan Wu, and Wei Liu.
\newblock Cnn in mrf: Video object segmentation via inference in a cnn-based
  higher-order spatio-temporal mrf.
\newblock In \emph{CVPR}, 2018.

\bibitem[Chen et~al.(2017{\natexlab{a}})Chen, Papandreou, Schroff, and
  Adam]{deeplabv3}
Liang-Chieh Chen, George Papandreou, Florian Schroff, and Hartwig Adam.
\newblock Rethinking atrous convolution for semantic image segmentation.
\newblock \emph{arXiv preprint arXiv:1706.05587}, 2017{\natexlab{a}}.

\bibitem[Chen et~al.(2017{\natexlab{b}})Chen, Liu, Li, Lu, and
  Song]{chen2017targeted}
Xinyun Chen, Chang Liu, Bo~Li, Kimberly Lu, and Dawn Song.
\newblock Targeted backdoor attacks on deep learning systems using data
  poisoning.
\newblock \emph{arXiv preprint arXiv:1712.05526}, 2017{\natexlab{b}}.

\bibitem[Choe et~al.(2020)Choe, Lee, and Shim]{choe2020attention}
Junsuk Choe, Seungho Lee, and Hyunjung Shim.
\newblock Attention-based dropout layer for weakly supervised single object
  localization and semantic segmentation.
\newblock \emph{IEEE Transactions on Pattern Analysis and Machine
  Intelligence}, 2020.

\bibitem[Feng et~al.(2020)Feng, Haase-Schuetz, Rosenbaum, Hertlein, Glaeser,
  Timm, Wiesbeck, and Dietmayer]{feng2020deep}
Di~Feng, Christian Haase-Schuetz, Lars Rosenbaum, Heinz Hertlein, Claudius
  Glaeser, Fabian Timm, Werner Wiesbeck, and Klaus Dietmayer.
\newblock Deep multi-modal object detection and semantic segmentation for
  autonomous driving: Datasets, methods, and challenges.
\newblock \emph{IEEE Transactions on Intelligent Transportation Systems}, 2020.

\bibitem[Fu et~al.(2019)Fu, Liu, Tian, Li, Bao, Fang, and Lu]{danet}
Jun Fu, Jing Liu, Haijie Tian, Yong Li, Yongjun Bao, Zhiwei Fang, and Hanqing
  Lu.
\newblock Dual attention network for scene segmentation.
\newblock In \emph{CVPR}, 2019.

\bibitem[Gu et~al.(2019)Gu, Liu, Dolan-Gavitt, and Garg]{gu2019badnets}
Tianyu Gu, Kang Liu, Brendan Dolan-Gavitt, and Siddharth Garg.
\newblock Badnets: Evaluating backdooring attacks on deep neural networks.
\newblock \emph{IEEE Access}, 7:\penalty0 47230--47244, 2019.

\bibitem[Guo et~al.(2020)Guo, Wang, Xu, Xing, Du, and Song]{guo2020towards}
Wenbo Guo, Lun Wang, Yan Xu, Xinyu Xing, Min Du, and Dawn Song.
\newblock Towards inspecting and eliminating trojan backdoors in deep neural
  networks.
\newblock In \emph{ICDM}, 2020.

\bibitem[Han et~al.(2020)Han, Zheng, Zhu, Xu, and Fang]{han2020live}
Lei Han, Tian Zheng, Yinheng Zhu, Lan Xu, and Lu~Fang.
\newblock Live semantic 3d perception for immersive augmented reality.
\newblock \emph{IEEE transactions on visualization and computer graphics},
  26\penalty0 (5):\penalty0 2012--2022, 2020.

\bibitem[Huang et~al.(2019)Huang, Wang, Huang, Huang, Wei, and
  Liu]{huang2019ccnet}
Zilong Huang, Xinggang Wang, Lichao Huang, Chang Huang, Yunchao Wei, and Wenyu
  Liu.
\newblock Ccnet: Criss-cross attention for semantic segmentation.
\newblock In \emph{ICCV}, 2019.

\bibitem[Li et~al.(2020{\natexlab{a}})Li, Wu, Jiang, Li, and
  Xia]{li2020backdoor1}
Yiming Li, Baoyuan Wu, Yong Jiang, Zhifeng Li, and Shu-Tao Xia.
\newblock Backdoor learning: A survey.
\newblock \emph{arXiv preprint arXiv:2007.08745}, 2020{\natexlab{a}}.

\bibitem[Li et~al.(2020{\natexlab{b}})Li, Li, Wu, Li, He, and
  Lyu]{li2020backdoor}
Yuezun Li, Yiming Li, Baoyuan Wu, Longkang Li, Ran He, and Siwei Lyu.
\newblock Backdoor attack with sample-specific triggers.
\newblock \emph{arXiv preprint arXiv:2012.03816}, 2020{\natexlab{b}}.

\bibitem[Liu et~al.(2020)Liu, Ma, Bailey, and Lu]{liu2020reflection}
Yunfei Liu, Xingjun Ma, James Bailey, and Feng Lu.
\newblock Reflection backdoor: A natural backdoor attack on deep neural
  networks.
\newblock In \emph{ECCV}, 2020.

\bibitem[Nguyen \& Tran(2021)Nguyen and Tran]{tuan2021wanet}
Anh~Tuan Nguyen and Tuan~Anh Tran.
\newblock Wanet - imperceptible warping-based backdoor attack.
\newblock In \emph{ICLR}, 2021.

\bibitem[Siam et~al.(2018)Siam, Gamal, Abdel-Razek, Yogamani, Jagersand, and
  Zhang]{siam2018comparative}
Mennatullah Siam, Mostafa Gamal, Moemen Abdel-Razek, Senthil Yogamani, Martin
  Jagersand, and Hong Zhang.
\newblock A comparative study of real-time semantic segmentation for autonomous
  driving.
\newblock In \emph{CVPR Workshop}, 2018.

\bibitem[Wang et~al.(2019)Wang, Yao, Shan, Li, Viswanath, Zheng, and
  Zhao]{wang2019neural}
Bolun Wang, Yuanshun Yao, Shawn Shan, Huiying Li, Bimal Viswanath, Haitao
  Zheng, and Ben~Y Zhao.
\newblock Neural cleanse: Identifying and mitigating backdoor attacks in neural
  networks.
\newblock In \emph{IEEE S\&P}, 2019.

\bibitem[Yang et~al.(2018)Yang, Yu, Zhang, Li, and Yang]{yang2018denseaspp}
Maoke Yang, Kun Yu, Chi Zhang, Zhiwei Li, and Kuiyuan Yang.
\newblock Denseaspp for semantic segmentation in street scenes.
\newblock In \emph{CVPR}, 2018.

\bibitem[Yao et~al.(2019)Yao, Li, Zheng, and Zhao]{yao2019latent}
Yuanshun Yao, Huiying Li, Haitao Zheng, and Ben~Y Zhao.
\newblock Latent backdoor attacks on deep neural networks.
\newblock In \emph{CCS}, 2019.

\bibitem[Zhai et~al.(2021)Zhai, Li, Zhang, Wu, Jiang, and
  Xia]{zhai2021backdoor}
Tongqing Zhai, Yiming Li, Ziqi Zhang, Baoyuan Wu, Yong Jiang, and Shu-Tao Xia.
\newblock Backdoor attack against speaker verification.
\newblock In \emph{ICASSP}, 2021.

\bibitem[Zhang et~al.(2019)Zhang, David, Foroosh, and
  Gong]{zhang2019curriculum}
Yang Zhang, Philip David, Hassan Foroosh, and Boqing Gong.
\newblock A curriculum domain adaptation approach to the semantic segmentation
  of urban scenes.
\newblock \emph{IEEE transactions on pattern analysis and machine
  intelligence}, 42\penalty0 (8):\penalty0 1823--1841, 2019.

\bibitem[Zhang et~al.(2020)Zhang, Zhou, David, Yue, Xi, Gong, and
  Foroosh]{zhang2020polarnet}
Yang Zhang, Zixiang Zhou, Philip David, Xiangyu Yue, Zerong Xi, Boqing Gong,
  and Hassan Foroosh.
\newblock Polarnet: An improved grid representation for online lidar point
  clouds semantic segmentation.
\newblock In \emph{CVPR}, 2020.

\bibitem[Zhou et~al.(2017)Zhou, Zhao, Puig, Fidler, Barriuso, and
  Torralba]{ade20k}
Bolei Zhou, Hang Zhao, Xavier Puig, Sanja Fidler, Adela Barriuso, and Antonio
  Torralba.
\newblock Scene parsing through ade20k dataset.
\newblock In \emph{CVPR}, 2017.

\bibitem[Zhu et~al.(2020)Zhu, Ning, Wang, Xin, and Wu]{zhu2020gangsweep}
Liuwan Zhu, Rui Ning, Cong Wang, Chunsheng Xin, and Hongyi Wu.
\newblock Gangsweep: Sweep out neural backdoors by gan.
\newblock In \emph{ACM MM}, 2020.

\end{thebibliography}
\bibliographystyle{iclr2021_conference}

\end{document}